\documentstyle[preprint,aps,epsfig]{revtex}
\tightenlines
\begin{document}

% \draft
\preprint{
\vbox{
% \hbox{August 2000}
\hbox{ADP-00-28/T411}
\hbox{JLAB-THY-00-17}
\hbox{PCCF RI 00-11}
\hbox{KSUCNR-107-00}
}}

\title{Neutron Structure Function and $A=3$ Mirror Nuclei}

\author{I.R.~Afnan$^1$, F.~Bissey$^{2,3}$, J.~Gomez$^4$,
        A.T.~Katramatou$^5$, W.~Melnitchouk$^{2,4}$,
        G.G.~Petratos$^5$, A.W.~Thomas$^2$}

\address{$^1$   School of Physical Sciences,
                Flinders University of South Australia,
                Bedford Park 5042,
                Australia}
\address{$^2$   Special Research Centre for the
                Subatomic Structure of Matter,
                and Department of Physics and Mathematical Physics,
                Adelaide University, Adelaide 5005,
                Australia}
\address{$^3$   Laboratoire de Physique Corpusculaire, 
                Universit\'e Blaise Pascal, CNRS/IN2P3,	\\
                24 avenue des Landais, 63177 Aubi\`ere Cedex,
                France}
\address{$^4$   Jefferson Lab,
                12000 Jefferson Avenue,
                Newport News, VA 23606}
\address{$^5$   Kent State University,
                Kent, OH 44242}

\maketitle

\begin{abstract}
We investigate deep inelastic scattering from $^3$He and $^3$H within a
conventional convolution treatment of binding and Fermi motion effects.
Using realistic Faddeev wave functions
together with a nucleon spectral function, %% ??
we demonstrate that the free
neutron structure function can be extracted in deep-inelastic scattering
from $A=3$ mirror nuclei, with nuclear effects canceling to within 2\%
for $x \alt 0.85$.
\end{abstract}

\newpage
%%%%%%%%%%%%%%%%%%%%%%%%%%%%%%%%%%%%%%%%%%%%%%%%%%%%%%%%%%%%%%%%%%%%%%%%%%%%%
One of the most fundamental properties of the nucleon is the structure
of its valence quark distributions.
Unlike the sea, which is generated via both perturbative and
non-perturbative mechanisms, the valence quark structure reflects
entirely large distance dynamics in the nucleon, which cannot be
described within perturbative quantum chromodynamics.

Experimentally, most of the recent studies of nucleon structure have
emphasized the small-$x$ region populated mainly by sea quarks ($x$
being the fraction of momentum of the nucleon carried by the quark),
while the valence quark structure has for some time now been thought
to be understood.
This is to some extent true, albeit with one major exception ---
the so-called deep valence region, at very large $x$, $x \agt 0.7$.
Recently it has become more widely appreciated that knowledge of quark
distributions at large $x$ is essential for a number of reasons.
Not least of these is the necessity of understanding backgrounds in
collider experiments, such as in searches for new physics beyond the
standard model \cite{CTEQ_LX}.
Furthermore, the behavior of the $d/u$ quark distribution ratio in the
limit $x \to 1$ is a critical test of the mechanism of spin-flavor
symmetry breaking in the nucleon, and of the onset of perturbative
behavior in large-$x$ structure functions.

The most widely used source of information about the valence quark
distributions in the nucleon has been the proton structure function,
$F_2^p$, which at large $x$ measures a charge-squared weighted
combination of the valence $u$ and $d$ distributions.
Because the $u$ quark is weighted by a factor 4:1 compared with the $d$,
the $F_2^p$ structure function most directly constrains the $u$ quark
distribution.

To determine the individual isospin distributions separately requires
a second linear combination of $u$ and $d$, which traditionally is
obtained from the neutron structure function, $F_2^n$, and which could
in principle constrain the $d$ quark distribution as well as the $u$.
However, the absence of free neutron targets means that usually the
deuteron is used as an effective neutron, with the neutron structure
function approximated by $F_2^n \approx F_2^d - F_2^p$.
While this approximation is valid at moderate $x$, it breaks down
dramatically for $x \agt 0.4$ due to Fermi motion and nuclear
binding effects in the deuteron \cite{WEST,OTHER,MSTD,MT}.

The problem of extracting neutron structure functions from nuclear data
is rather old \cite{WEST}, although recently the discussion has been
revived with the realization \cite{MT} that $F_2^n$, extracted from
$F_2^d$ by taking into account Fermi motion and binding (off-shell)
effects, could be significantly larger than that extracted in earlier
analyses in which only Fermi motion corrections were applied.
In particular, omitting nuclear binding corrections can introduce errors
of up to 50\%~\cite{MT,WHITLOW} in $F_2^n/F_2^p$ already at $x \sim 0.75$.
Such a difference is of the same order of magnitude as the variation of
the behavior of the $F_2^n/F_2^p$ ratio predicted in the $x \to 1$ limit,
which ranges from 1/4 in non-perturbative models where the $d$ quark is
suppressed relative to $u$ \cite{QUARTER}, to 3/7 in perturbative
QCD-inspired models which emphasize helicity aligned configurations of
the quark and nucleon \cite{FJ}.

Although one can make a strong argument that a proper treatment of
nuclear corrections in the deuteron should account for both Fermi
motion as well as binding effects, the question can ultimately be
settled only by experiment.
In this paper we suggest how this can be achieved by using a novel method
which maximally exploits the mirror symmetry of $A=3$ nuclei.
Regardless of the absolute value of the nuclear EMC effects in $^3$He or
$^3$H, the differences between these will be small --- on the scale of
charge symmetry breaking in the nucleus --- which allows a relatively
clean determination of $F_2^n$ over a large range of $x$ essentially free
of nuclear contamination.

The argument is actually rather simple.
In the absence of the Coulomb interaction and in an isospin symmetric
world the properties of a proton (neutron) bound in a $^3$He nucleus
would be identical to that of a neutron (proton) bound in $^3$H.
If, in addition, the proton and neutron distributions in $^3$He
(and in $^3$H) were identical, the neutron structure function could
be extracted with no nuclear corrections, regardless of the
size of the EMC effect in $^3$He or $^3$H separately.

In practice, $^3$He and $^3$H are of course not perfect mirror nuclei
--- their binding energies for instance differ by some 10\% --- and the
$p$ and $n$ distributions are not quite identical.
However, the $A=3$ system has been studied for many years, and modern
realistic $A=3$ wave functions are known to rather good accuracy.
In a self-consistent framework one can use the same $NN$ interaction
which describes the two-nucleon system (e.g. $NN$ scattering, deuteron
form factors, quasi-elastic $e d$ scattering) to provide the basic input
interaction into the three-nucleon calculation.
Therefore the wave functions can be tested against a large array of
observables which put rather strong constraints on the models.

Defining the EMC-type ratios for the $F_2$ structure functions of
$^3$He and $^3$H (weighted by corresponding isospin factors) by:
\begin{mathletters}
\begin{eqnarray}
R(^3{\rm He}) &=& { F_2^{^3{\rm He}} \over 2 F_2^p + F_2^n }\ , \\
R(^3{\rm H}) &=& { F_2^{^3{\rm H}} \over F_2^p + 2 F_2^n }\ ,
\end{eqnarray}  
\end{mathletters}%
one can write the ratio of these as:
\begin{eqnarray}
\label{rr}
{\cal R} &=& { R(^3{\rm He}) \over R(^3{\rm H}) }\ .
\end{eqnarray}
Inverting this expression directly yields the ratio of the free neutron
to proton structure functions:
\begin{eqnarray}
\label{np}
{ F_2^n \over F_2^p }
&=& { 2 {\cal R} - F_2^{^3{\rm He}}/F_2^{^3{\rm H}}
\over 2 F_2^{^3{\rm He}}/F_2^{^3{\rm H}} - {\cal R} }\ .
\end{eqnarray}
We stress that $F_2^n/F_2^p$ extracted via Eq.(\ref{np}) does not depend
on the size of the EMC effect in $^3$He or $^3$H, but rather on the
{\em ratio} of the EMC effects in $^3$He and $^3$H.
If the neutron and proton distributions in the $A=3$ nuclei are not
dramatically different, one might expect ${\cal R} \approx 1$.
To test whether this is indeed the case requires an explicit calculation
of the EMC effect in the $A=3$ system.

The conventional approach employed in calculating nuclear structure 
functions in the valence quark region, $x \agt 0.3$, is the impulse
approximation, in which the virtual photon scatters incoherently from
individual nucleons in the nucleus \cite{CONV}.
The nuclear cross section is determined by factorizing the
$\gamma^*$--nucleus interaction into $\gamma^*$--nucleon and
nucleon--nucleus amplitudes.
In the absence of relativistic and nucleon off-shell corrections
\cite{MST,OFFSHELL}, the structure function of a nucleus, $F_2^A$,
can then be calculated by folding the nucleon structure function,
$F_2^N$, with a nucleon momentum distribution in the nucleus, $f_{N/A}$:
\begin{eqnarray}
\label{con}
F_2^A(x) &=& \int dy\ f(y)\ F_2^N(x/y)\
\equiv\ f(x) \otimes F_2^N(x)\ ,
\end{eqnarray}
where $y$ is the fraction of the `plus'-component of the nuclear momentum
carried by the interacting nucleon, and the $Q^2$ dependence in the
structure functions is implicit.
The convolution expression in Eq.(\ref{con}) is correct in the limit of
large $Q^2$; at finite $Q^2$ there are additional contributions to
$F_2^A$ from the nucleon $F_1^N$ structure function, although these are
suppressed by powers of $M^2/Q^2$, where $M$ is the nucleon mass.
Corrections to the impulse approximation appear in the guise of final
state interactions, multiple rescattering (nuclear shadowing), $NN$
correlations and 6-quark clusters, however, these are generally confined
to either the small-$x$ \cite{SHAD}, or very large-$x$ ($x \agt 0.9$)
\cite{NNCOR} regions.

The distribution $f(y)$ of nucleons in the nucleus is related to the
nucleon spectral function $S(p)$ by \cite{CONV}:
\begin{eqnarray}
f(y)
&=& \int d^4 p\
\left( 1 + {p_z \over p_0} \right)
\delta \left( y - { p_0 + p_z \over M } \right) S(p)\ ,
\end{eqnarray}
where $p$ is the four-momentum of the bound nucleon, and is normalized
such that $\int dy\ f(y) = 1$.
The spectral function includes final state interactions between the
two spectator nucleons, either as a deuteron or in the continuum.
In the latter case, this includes an integration over interacting final
state $NN$ wave functions, as distinct from a calculation in terms of
a simple momentum distribution (see Ref.\cite{SPECFN} for a definition
of the three-body spectral function).
For an $A=3$ nucleus the spectral function is evaluated from the
three-body nuclear wave function, calculated by solving the homogeneous
Faddeev equation with a given two-body interaction.
Details of the computation of the wave functions can be found in
Ref.\cite{BTA}.
To examine the model dependence of the distribution function we use
several different potentials, namely the ``EST'' (Ernst-Shakin-Thaler)
separable approximation to the Paris potential \cite{PEST} (referred to
as ``PEST''), the unitary pole approximation \cite{SA} to the Reid Soft
Core (RSC) potential, and the Yamaguchi potential \cite{YAM} with 7\%
mixing between $^3 S_1$ and $^3 D_1$ waves.

In terms of the proton and neutron momentum distributions, the $F_2$
structure function for $^3$He is given by:
\begin{mathletters}
\begin{eqnarray}
F_2^{^3{\rm He}}\
&=& 2\ f_{p/^3{\rm He}}\ \otimes\ F_2^p\
 +\    f_{n/^3{\rm He}}\ \otimes\ F_2^n\ .
\end{eqnarray}
Similarly for $^3$H, the structure function is evaluated from the proton
and neutron momentum distributions in $^3$H:
\begin{eqnarray}
F_2^{^3{\rm H}}
&=&    f_{p/^3{\rm H}} \otimes\ F_2^p\
 +\ 2\ f_{n/^3{\rm H}} \otimes\ F_2^n\ .
\end{eqnarray}
\end{mathletters}%
Because isospin symmetry breaking effects in nuclei are quite small,
one can to a good approximation relate the proton and neutron
distributions in $^3$He to those in $^3$H:
\begin{mathletters}
\begin{eqnarray}
f_{n/^3{\rm H}} &\approx& f_{p/^3{\rm He}}\ ,   \\
f_{p/^3{\rm H}} &\approx& f_{n/^3{\rm He}}\ ,
\end{eqnarray}
\end{mathletters}%
although in practice we consider both the isospin symmetric and isospin
symmetry breaking cases explicitly.
Note that even in the isospin symmetric case the proton and neutron
distributions in $^3$He will be different because while the neutron in
$^3$He is accompanied by a spectator $pp$, the spectator system of the 
proton is either an uncorrelated $pn$ pair or a recoiling deuteron.

The ratio ${\cal R}$ of EMC ratios for $^3$He and $^3$H is shown in
Fig.~1 for the various nuclear model wave functions (PEST, RSC and
Yamaguchi), using the CTEQ parameterization \cite{CTEQ} of parton
distributions at $Q^2=10$~GeV$^2$ for $F_2^N$.
The EMC effects are seen to largely cancel over a large range of $x$,
out to $x \sim 0.85-0.9$, with the deviation from a `central value'
${\cal R} \approx 1.01$ within $\pm 1\%$.
Furthermore, the dependence on the nuclear wave function is very weak.
In practice, the exact shape of ${\cal R}$ will not be important for
the purposes of extracting $F_2^n/F_2^p$ from the
$F_2^{^3{\rm He}}/F_2^{^3{\rm H}}$ ratio; rather, it is essential that,
as we find, the model dependent deviation of ${\cal R}$ from the central
value should be small.

The dependence of ${\cal R}$ on the input nucleon structure function
parameterization is illustrated in Fig.~2, where several representative
curves at $Q^2 = 10$ GeV$^2$ are given: apart from the standard CTEQ
fit (solid), the results for the GRV \cite{GRV} (dot-dashed),
Donnachie-Landshoff (DL) \cite{DOLA} (dashed),
and BBS \cite{BBS} (dotted) parameterizations are also shown
(the latter at $Q^2=4$~GeV$^2$).
For $x \alt 0.6$ there is little dependence ($\alt 0.5\%$) in the
ratio on the structure function input.
For $0.6 \alt x \alt 0.85$ the dependence is greater, but still
with $\alt \pm 1\%$ deviation away from the central value
${\cal R} \approx 1.01$.
The spread in this region is due mainly to the poor knowledge of the
neutron structure function at large $x$.
Beyond $x \approx 0.85$ there are few data in the deep-inelastic region
on either the neutron or proton structure functions, so here both the $d$
and $u$ quark distributions are poorly determined.

A standard assumption in most global fits of parton distributions is
that $d/u \to 0$ as $x \to 1$.
This assumption has recently been questioned on theoretical and
phenomenological grounds \cite{MT,W,DU}.
The BBS parameterization \cite{BBS}, on the other hand, incorporates
constraints from perturbative QCD, and forces $d/u \to 0.2$ as $x \to 1$
\cite{FJ}.
The effect of the different large-$x$ behavior of the $d$ quark is
apparent only for $x \agt 0.85$, where it gives a difference of
$\sim$ 1--2\% in ${\cal R}$ compared with the fits in which $d/u \to 0$.
One can also modify the standard CTEQ fit, for example, by applying a
correction factor \cite{W} to enforce $d/u \to 0.2$, however, this also
produces differences in ${\cal R}$ which are $\alt 2\%$ for $x < 0.9$.

Despite the seemingly strong dependence on the nucleon structure function
input at very large $x$, this dependence is actually artificial.
In practice, once the ratio $F_2^{^3{\rm He}}/F_2^{^3{\rm H}}$ is
measured, one can employ an iterative procedure to eliminate this
dependence altogether.
Namely, after extracting $F_2^n/F_2^p$ from the data using some
calculated ${\cal R}$, the extracted $F_2^n$ can then be used to
compute a new ${\cal R}$, which is then used to extract a new and
better value of $F_2^n/F_2^p$.
This procedure is iterated until convergence is achieved and a
self-consistent solution for the extracted $F_2^n/F_2^p$ and
${\cal R}$ is obtained (see also Ref.\cite{PSS}).

All of the structure functions discussed thus far have been calculated
assuming leading twist dominance at $Q^2=10$~GeV$^2$.
To test the sensitivity of the ratio to possible effects beyond leading
twist, we have calculated ${\cal R}$ using the fit to the total $F_2$
structure function from Donnachie and Landshoff \cite{DOLA}, which has an
explicit higher twist ($\propto 1/Q^2$) component in addition to the
leading twist.
The result is indicated by the upper dot-dashed curve DL(HT) in Fig.~2.
The difference between the leading twist only and leading $+$ higher
twist curves is negligible for $x \alt 0.8$, increasing to $\sim 1.5\%$
at $x \sim 0.85$, where higher twist effects are known to be more
important.
The size of the higher twist corrections can be determined by taking
measurements at several values of $Q^2$ and observing any $1/Q^2$
dependence of the structure function.
In particular, since the $Q^2$ dependence of $F_2^p$ has been measured
in a number of earlier experiments \cite{HT}, the $Q^2$ dependence of
the extracted $F_2^n/F_2^p$ can be used to separate the leading twist
from the non-leading twist components of $F_2^n$.

We conclude therefore that the effect on ${\cal R}$ from the present
lack of knowledge of the nucleon structure function is $\alt 2\%$
for $x \alt 0.85$.
However, this uncertainty can in principle be eliminated altogether
via an iteration procedure, so that the only model dependence of
${\cal R}$ will be from the nuclear interaction in the $A=3$ nucleus.

The ratios in Fig.~1 were calculated using three-nucleon wave functions
neglecting the Coulomb interaction and working in an isospin basis
(we also omit possible three-body forces since these are expected to
have a negligible effect on ${\cal R}$).
To estimate the effect of neglecting the Coulomb interaction in $^3$He
and at the same time correct the long range part of the three-body wave
function due to the change in the binding energy, we have modified the
$^1S_0$ potential in $^3$He and $^3$H to reproduce their respective
experimental binding energies.
In this way the $^3S_1-^3D_1$ interaction responsible for the formation
of the deuteron is unchanged.
This approximation spreads the effect of the Coulomb interaction over
both the $pp$ and $np$ interaction in the $^1S_0$ channel.
To that extent, it shifts some of the Coulomb effects in the neutron 
distribution in $^3$He to the proton distribution.
However, this simple modification to the $^1S_0$ interaction will allow
us to study explicitly the possible effects associated with the
differences in the binding energies of $^3$He and $^3$H.

The ratio ${\cal R}$ calculated with the PEST wave function modified
according to this prescription is shown in Fig.~3, labeled PEST(E)
(dashed curve).
(The CTEQ parameterization of the nucleon structure function at
$Q^2=10$~GeV$^2$ is used.)
The result of this modification is a shift of approximately 0.5--1\%
shift in ${\cal R}$, with the net effect still being a ratio which
deviates by $< 2\%$ from unity.

Also shown in Fig.~3 is the prediction of the nuclear density model,
extrapolated from heavy nuclei to $A=3$ \cite{FS88}.
The nuclear density model, which has proven successful for studying
the $A$-dependence of the EMC effect for heavy nuclei, stems from the 
empirical observation that for heavy nuclei the deviation from unity
in the EMC ratio is assumed to scale with nuclear density:
\begin{eqnarray}
\label{dens}
{ R(A_1)-1 \over R(A_2)-1 }
&=& { \rho(A_1) \over \rho(A_2) }\ ,
\end{eqnarray}
where $\rho(A)$ is the mean nuclear density.
{}From the empirical $A=3$ charge radii one finds that
$\rho(^3{\rm H})/\rho(^3{\rm He}) \approx 140\%$, so that the EMC effect
in $^3$H is predicted to be $\sim 40\%$ bigger than in $^3$He.
However, assuming that $R(^3{\rm He})$ can be extrapolated from the
measured EMC ratios for heavy nuclei such as $^{56}$Fe, one still finds
that ratio $|{\cal R}-1| < 2\%$ for all $x \alt 0.85$.
Although there are questions about the meaning of nuclear density for
a few-body system \cite{MABT}, it is reassuring to see that practically
the entire range of models of the nuclear EMC effect predict that
${\cal R}$ is within 1--2\% of unity for all $x \alt 0.85$.

The ideal place to carry out a high-$x$ deep-inelastic scattering (DIS)
experiment on $^3$He and $^3$H \cite{WKSHP_GGP,WKSHP_WM} is Jefferson Lab
(JLab) with its proposed energy upgrade to 12 GeV.
Since the ratio of longitudinal to transverse photoabsorption cross
sections $R=\sigma_L/\sigma_T$ is the same for $^3$He and $^3$H,
measurements of the $^3$He and $^3$H DIS cross sections under identical
conditions can provide a direct measurement of the ratio of the
$F_2$ structure functions of the two nuclei:
$\sigma({\rm ^3H}) / \sigma({\rm ^3He}) =
F_2^{\rm ^3H} / F_2^{\rm ^3He}$.
The key issue for the experiment will be the availability of a high
density $^3$H tritium target similar to those used in the past to measure
the elastic form factors of $^3$H at Saclay \cite{SACLAY} and MIT-Bates
\cite{BATES}.
The high intensity of the JLab beam and the large acceptance of existing
or proposed JLab spectrometers will facilitate high statistics DIS cross
section measurements ($\le \pm 0.25 \%$) over a large $x$ range
($0.10 \leq x \leq 0.83$) and valuable systematics checks in a data
taking period of just a few weeks.

The measured $F_2^{^3\rm H}/F_2^{^3\rm He}$ ratio is expected to be
dominated by experimental uncertainties that do not cancel in the DIS
cross section ratio of $^3$H to $^3$He, and the theoretical uncertainty
in the calculation of ${\cal R}$.
Assuming that the target densities can be known to the $\simeq 0.5\%$
level and that the relative difference in the $^3$H and $^3$He radiative
corrections would be $\simeq 0.5\%$, the total experimental error in the
DIS cross section ratio of $^3$H to $^3$He should be $\le 1.0\%$ (similar
to the error of past DIS measurements of the proton to deuteron cross
section ratio \cite{BODEK}).
Such an error is comparable to the present theoretical uncertainty in the
calculation of the ratio ${\cal R}$.

Figure 4 shows the presently available data on $F_2^n/F_2^p$,
adjusted for the JLab 12 GeV kinematics, as extracted from the SLAC
deep-inelastic $\sigma(p)$ and $\sigma(d)$ cross sections using a
Fermi-smearing model with the Paris nucleon--nucleon potential
\cite{WHITLOW}.
To indicate the quality of the proposed $F_2^n/F_2^p$ ratio determination
from the $\sigma(^3\rm H)/\sigma(^3\rm He)$ measurement, we plot in Fig.~4
the $\pm$ one standard deviation projected error band for the $x$ range
accessible with a 12 GeV upgraded JLab beam.
The band includes both projected experimental and theoretical
uncertainties.
The central values of the band represent $F_2^n/F_2^p$ determined using
the density model \cite{FS88} for the nuclear EMC effect and data on EMC
ratios for heavy nuclei from the SLAC experiment E139 \cite{GOMEZ}.
It is evident, therefore, that the proposed measurement will be able
to unambiguously distinguish between the two different methods of
extracting the $F_2^n/F_2^p$ ratio from proton and deuterium DIS
measurements, and determine its value for large $x$ with an excellent
precision in an (almost) independent model way.

As well as offering a relatively clean way to extract $F_2^n/F_2^p$,
DIS from the $^3$He/$^3$H system can also determine the absolute size of
the EMC effect in $A \leq 3$ nuclei.
With $F_2^n$ determined from the combined
$F_2^{^3{\rm He}}/F_2^{^3{\rm H}}$ and $F_2^p$ structure functions,
the size of the EMC effect in the deuteron (namely, $F_2^d/(F_2^p+F_2^n)$)
can be deduced from the measured $F_2^d/F_2^p$ ratio.
This would settle a question which has remained controversial since
the early 1970s.
Furthermore, data on the absolute values of $F_2^{^3{\rm He}}$ and
$F_2^{^3{\rm H}}$ will also allow the absolute value of the EMC effect
in $A=3$ nuclei to be determined.
To date the only data on $F_2^{^3{\rm He}}$ in existence are those
from the HERMES experiment \cite{HERMES}, which measured the ratio
$\sigma(^3{\rm He})/(\sigma(d)+\sigma(p))$, although the focus there
was the region of small $x$ and $Q^2$.

In summary, we have demonstrated the effectiveness of using $A=3$ mirror
nuclei to extract the ratio of the neutron to proton structure functions,
$F_2^n/F_2^p$, essentially free of nuclear effects for all $x \alt 0.85$.
A successful program of DIS measurements of $A=3$ cross sections at an
energy-upgraded Jefferson Lab would not only settle a ``text-book'' issue
which has eluded a definitive resolution for nearly 30 years, but would
also allow the completion of the empirical study of nuclear effects in
deep-inelastic scattering over the full range of mass numbers.

%%%%%%%%%%%%%%%%%%%%%%%%%%%%%%%%%%%%%%%%%%%%%%%%%%%%%%%%%%%%%%%%%%%%%%%%%%
\acknowledgements

We would like to thank S.~Liuti, G.~Salme, S.~Scopetta and S.~Simula
for helpful discussions and communications.
This work was supported by the Australian Research Council, the US
Department of Energy contract DE-AC05-84ER40150, and the US National
Science Foundation grant PHY-9722640.

%%%%%%%%%%%%%%%%%%%%%%%%%%%%%%%%%%%%%%%%%%%%%%%%%%%%%%%%%%%%%%%%%%%%%%%%%%

%%%%%%%%%%%%%%%%%%%%%%%%%%%%%%%%%%%%%%%%%%%%%%%%%%%%%%%%%%%%%%%%%%%%%%%%%%
\begin{figure}
\begin{center}
\epsfig{figure=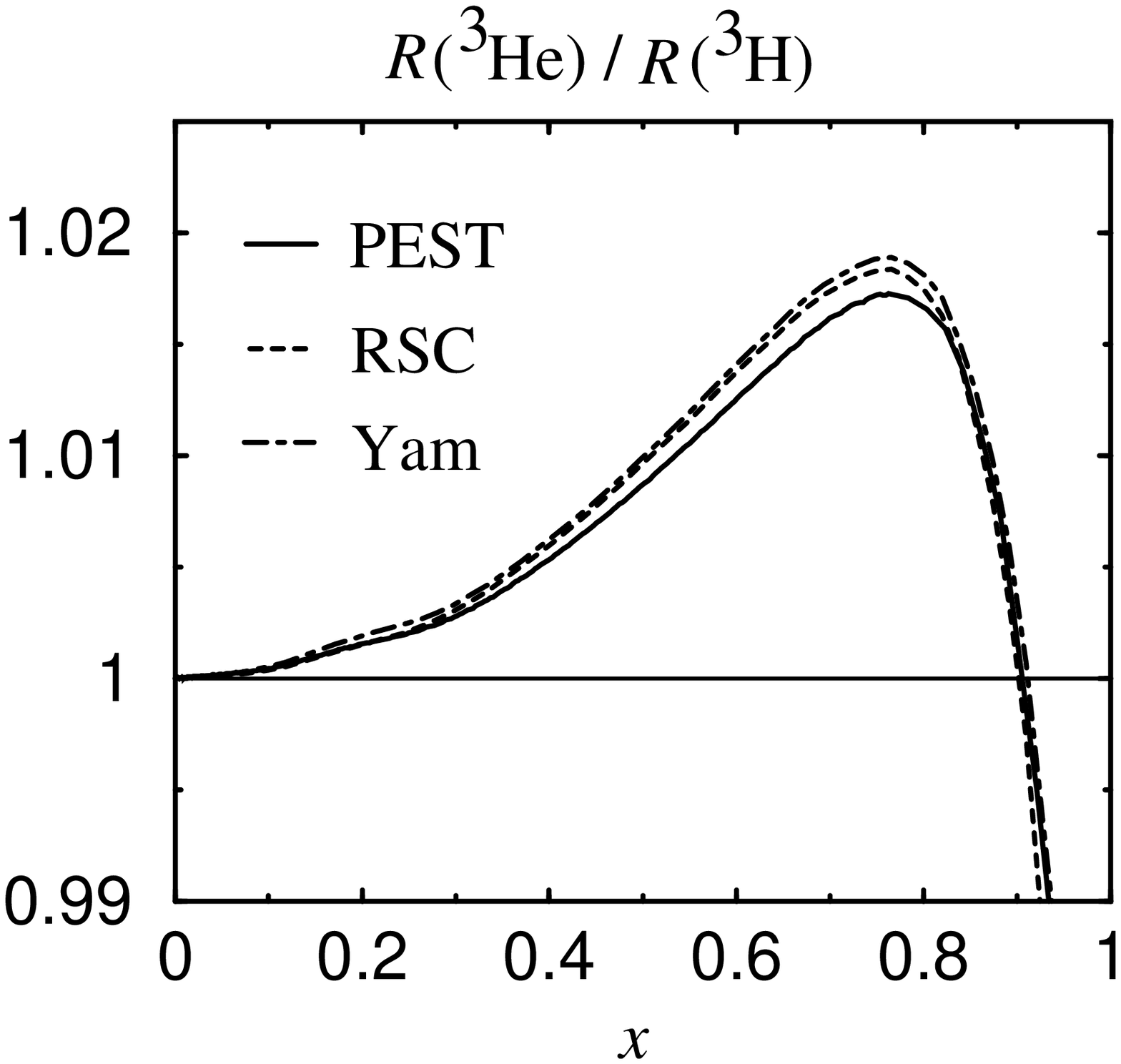,height=13cm}
\vspace*{1cm}
\caption{Ratio of nuclear EMC ratios for $^3$He and $^3$H for
        various nuclear models:
        PEST (solid),
        Reid Soft Core (dashed),
        Yamaguchi (dot-dashed).}
\end{center}
\end{figure}

\begin{figure}
\begin{center}
\epsfig{figure=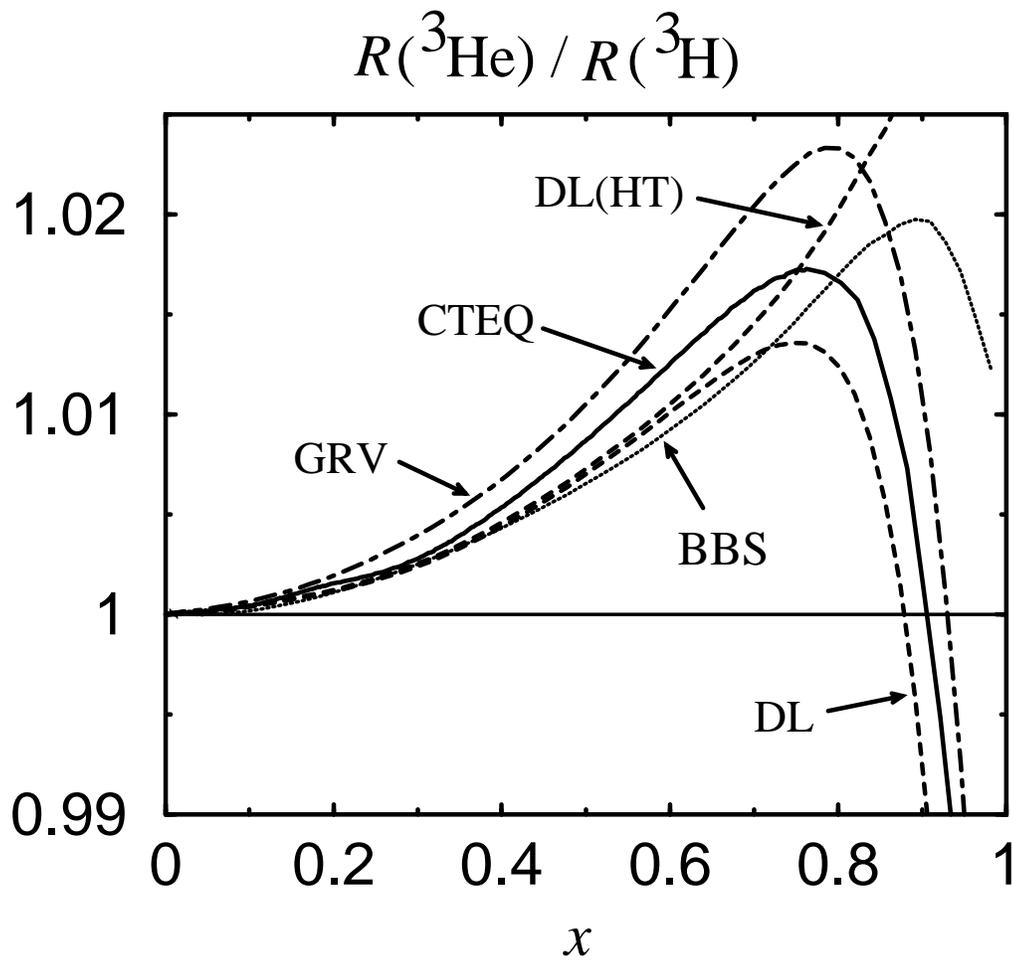,height=13cm}
\vspace*{1cm}
\caption{Ratio of nuclear EMC ratios for $^3$He and $^3$H with the
        PEST wave functions, using various nucleon structure function
        parameterizations:
        CTEQ (solid),
        GRV (dot-dashed),
        BBS (dotted),
        and Donnachie-Landshoff (DL) with leading twist only,
        and with higher twist (HT) correction (dot-dashed).}
\end{center}
\end{figure}

\begin{figure}
\begin{center}
\epsfig{figure=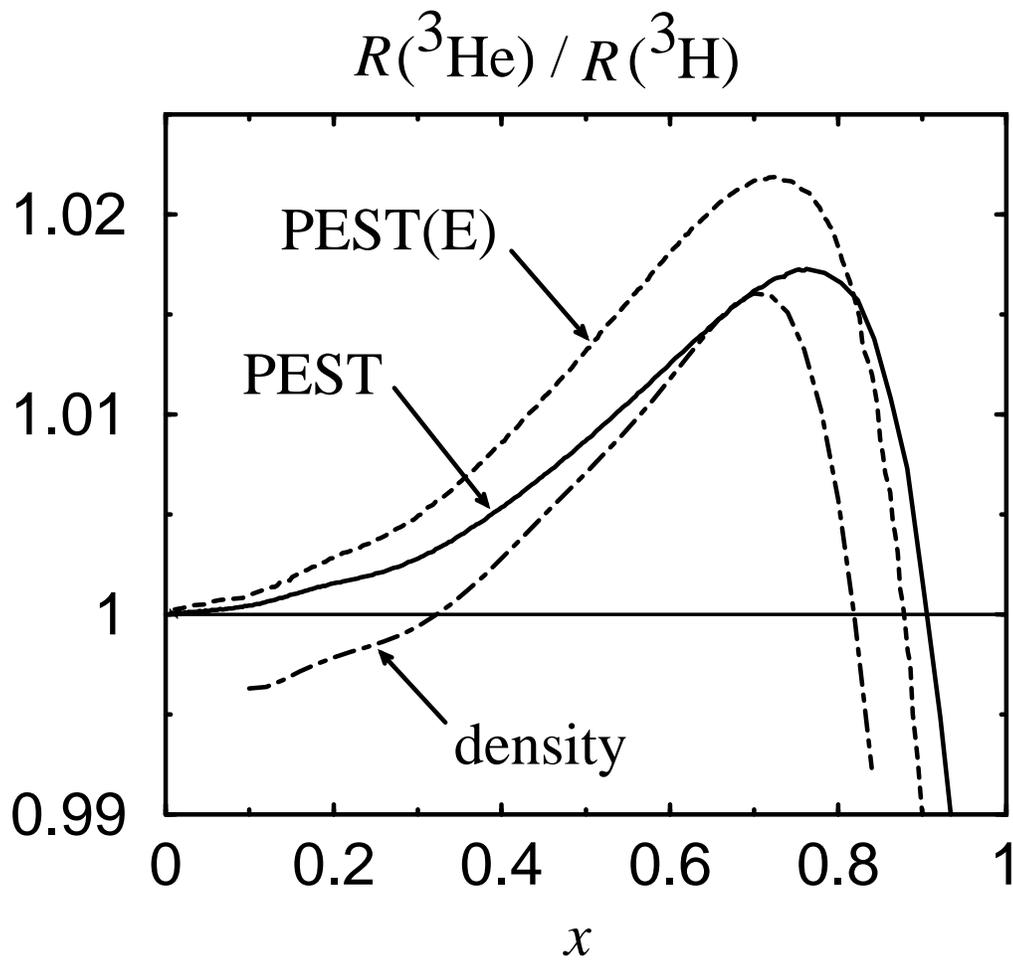,height=13cm}
\vspace*{1cm}
\caption{Ratio of nuclear EMC ratios for $^3$He and $^3$H for the
        PEST wave function (solid), modified PEST to reproduce the
        experimental binding energies (dashed), and the density
        extrapolation model (dot-dashed).}
\end{center}
\end{figure}

\begin{figure}
\begin{center}
\epsfig{figure=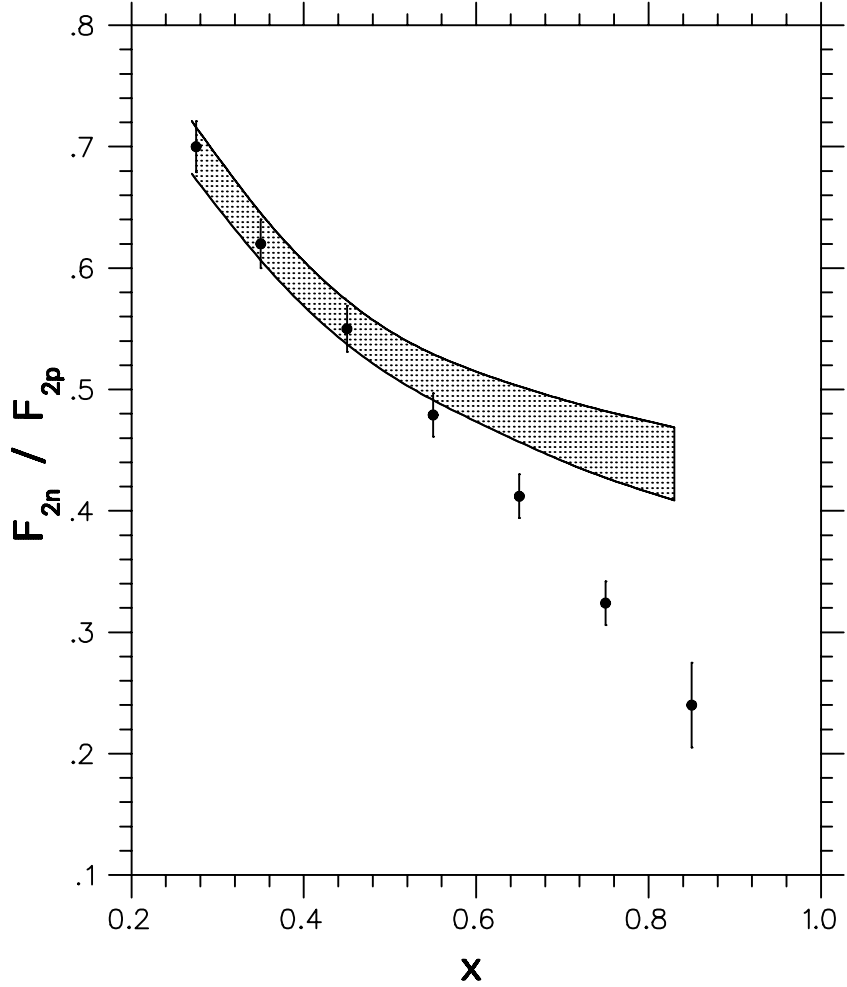,height=13cm}
\vspace*{1cm}
\caption{$F_2^n/F_2^p$ ratio extracted from previous deep-inelastic
	$p$ and $d$ cross sections using a Fermi-smearing
	model~\protect\cite{WHITLOW} (solid circles).
	The shaded band represents a $\pm$ one standard deviation error
	for the proposed $^3$H and $^3$He DIS JLab experiment, with the
	central values of the band corresponding to $F_2^n/F_2^p$
	extracted assuming an EMC effect in deuterium (see text).}
\end{center}
\end{figure}

\end{document}